\documentclass[twocolumn,preprintnumbers,amsmath,amssymb]{revtex4}

\usepackage{graphicx}
\usepackage{dcolumn}
\usepackage{bm}
\usepackage{color}
\usepackage{ulem}

\begin{document}%

%
%


\title{
Successive magnetic transitions in heavy fermion superconductor Ce$_{3}$PtIn$_{11}$ \\
studied by $^{115}$In nuclear quadrupole resonance
}

\author{
Hideto Fukazawa~\footnote{hideto@chiba-u.jp}$^{1}$, Kazuki Kumeda$^{1}$, Naoki Shioda$^{1}$, Yongsun Lee$^{1}$, Yoh Kohori$^{1}$, \\
Koudai Sugimoto$^{2}$, Debarchan Das$^{3,4}$, Joannna B\l awat$^{3}$ and Dariusz Kaczorowski$^{3}$
}
\address{
$^1$Department of Physics, Chiba University, Chiba 263-8522, Japan \\
$^2$Department of Physics, Keio University, Yokohama 223-8522, Japan\\
$^3$Institute of Low Temperature and Structure Research, Polish Academy of Sciences, P.O. Box 1410, PL-50-950 Wroc{\l}aw, Poland\\
$^4$ Laboratory for Muon Spin Spectroscopy, Paul Scherrer Institute, CH-5232 Villigen PSI, Switzerland
}
%
\date{October 13, 2020}
\begin{abstract}
Nuclear quadrupole resonance (NQR) measurements were performed on the heavy fermion superconductor Ce$_{3}$PtIn$_{11}$ with $T_{\rm c} = 0.32$~K. 
The temperature dependence of both spin-lattice relaxation rate $1/T_{1}$ and NQR spectra evidences the occurrence of two successive magnetic transitions 
with $T_{\rm N1} \simeq 2.2$~K and $T_{\rm N2} \simeq 2.0$~K. 
In successive magnetic transitions, even though the magnetic moment at the Ce(2) site plays a major role,  the magnetic moment at the Ce(1) site also  contributes to some extent. 
While a commensurate antiferromagnetic ordered state appears for $T_{\rm N2} < T < T_{\rm N1}$, a partially incommensurate antiferromagnetic ordered state is suggested for $T<T_{\rm N2}$.
\end{abstract}


\maketitle

\section{Introduction}

Ce-based intermetallic systems have gained considerable research interest due to their diverse and perplexing physical properties 
encompassing observation of non-Fermi-liquid behavior, heavy-fermion-superconductivity, and quantum criticality~\cite{Ste1,Math,Pet1,Wan1,Das1}. 
Among the vast number of known compounds, the members of the Ce$_{n}M_{m}$In$_{3n+2m}$ ($M=$~Co, Rh, Pd, Ir, Pt) 
family have drawn an unprecedented amount of research effort due to their fascinating phase diagrams, 
which allow exploration of the relationship between magnetism and adjacent superconductivity. 

Among them, CeIn$_{3}$ ($m=0$, $n = 1$) is a three-dimensional material, 
and the localized magnetism of Ce ions is dominant at ambient pressure ($T_{\rm N} = 10$~K) ~\cite{Math}. 
However, with increasing pressure, antiferromagnetic order is gradually suppressed, 
the hybridization between Ce $f$ electrons and conduction electrons becomes significant, 
and finally superconductivity emerges near the magnetic quantum critical point ($T_{\rm c} = 0.2$~K at 2.6~GPa). 
In turn, the Ce$M$In$_{5}$ phases ($m = n = 1$) have quasi-two-dimensional character. 
CeCoIn$_{5}$ is one of the most intensively studied systems in recent times due to its unique physics 
associated with heavy-fermion superconductivity at ambient pressure 
($T_{\rm c} = 2.3$~K) and normal-state non-Fermi-liquid behavior~\cite{Pet1,Koh1}. 
In high magnetic fields, confined in the basal plane of the tetragonal unit cell, an inhomogeneous superconducting phase of 
the Fulde-Ferrell-Larkin-Ovchinnikov (FFLO) type possibly forms at the lowest temperatures~\cite{Rad1,Bia1,Yana1}.
The compound CeRhIn$_{5}$, isostructural to CeCoIn$_{5}$, is antiferromagnetic at ambient pressure ($T_{\rm N} = 3.8$~K), and 
exhibits superconductivity under pressure ($T_{\rm c} = 2.1$~K at 2.0~GPa)~\cite{Heg1,Kaw1}. 
Maximum $T_{\rm c}$ is about an order of magnitude higher than that of CeIn$_{3}$, 
which is a consequence of the reduced effective dimensionality~\cite{Mon1}.
Ce$_{2}$PdIn$_{8}$ ($m=1$, $n = 2$) exhibits superconductivity below $T_{\rm c} = 0.64$~K~\cite{Kac1,Kac3}. 
There, the antiferromagnetic fluctuations are not as strong as in CeCoIn$_{5}$~\cite{Fuk4,Fuk5}. 

\begin{figure}
\includegraphics[width=8.5cm]{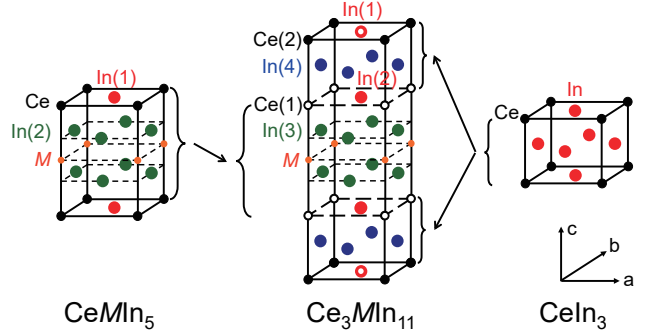}
\caption{
Crystal structures of the Ce$_{n}M$In$_{3n+2}$ ($n = 1,3,\infty,\, M=$~Co, Rh, Pd, Ir, Pt) compounds. 
One of the principal axes of the electric field gradient $V_{zz}$ is 
perpendicular to the crystal $ab$ plane for the In(1) and In(2) sites and parallel to the $a$ or $b$ axis for the In(3) and In(4) sites~\cite{Kam1,Kam2}. 
}
\label{f1}
\end{figure}

Recently, the synthesis of Ce$_{3}M$In$_{11}$ ($m=1$, $n = 3$) has been reported~\cite{Kac4,Cus1}. 
Unlike Ce$M$In$_{5}$, Ce$_{2}M$In$_{8}$, and CeIn$_{3}$, in which a single Ce site exists, 
there are two inequivalent Ce sites in the unit cell of the Ce$_{3}M$In$_{11}$ crystal, namely, Ce(1) at the $4mm$ site and Ce(2) at the $4/mmm$ site. 
The structure gives a new interest in considering magnetism and superconductivity in the Ce$_{n}M_{m}$In$_{3n+2m}$ group. 
Both Ce$_{3}$PtIn$_{11}$ and Ce$_{3}$PdIn$_{11}$ exhibit two successive antiferromagnetic phase transitions 
at ambient pressure ($T_{\rm N1} \simeq 2.2$~K and $T_{\rm N2} \simeq 2.0$~K for $M =$~Pt, 
$T_{\rm N1} \simeq 1.7$~K and $T_{\rm N2} \simeq 1.5$~K for $M =$~Pd)~\cite{Cus1}. 
Even in the antiferromagnetic state, they exhibit superconductivity below $T_{\rm c} = 0.32$~K ($M =$~Pt) and 0.42~K ($M =$~Pd)~\cite{Cus2,Cus3}. 
In each material, a further metamagnetic transition occurs 
in a high magnetic field~\cite{Kac5,Das2}.
This suggests that the coexistence of antiferromagnetism and superconductivity in the Ce$_{3}M$In$_{11}$ compounds is quite a unique phenomenon 
driven by the hybridization between $f$electrons of Ce ions and conduction electrons~\cite{Ben1}. 
Naively, the superconductivity and two successive antiferromagnetic transitions can be considered to be due to two inequivalent Ce sites.  
The localized magnetic moment of the Ce(2) site (depicted in Fig.\hspace{-2pt}\ref{f1}) is supposed to be responsible for the magnetic transitions. 
In turn, the Ce(1) site is either nonmagnetic or very weakly magnetic, 
as suggested by the specific heat data~\cite{Cus1,Cus2}. 
This duality of Ce sites facilitates understanding of new physics near the quantum critical point. 
For this purpose, knowledge of magnetic structures in the Ce$_{3}M$In$_{11}$ compounds is essential. 
However, due to excessive In content in this system, standard neutron diffraction measurements are 
difficult to perform and time consuming because In has a large absorption cross section. 
Thus local probe measurements such as nuclear quadrupole resonance (NQR) can be of great use,
and indeed such experiments have recently been performed for Ce$_{3}$PtIn$_{11}$ by Kambe {\it et al.}~\cite{Kam1,Kam2}. 
However, it must be noted that the successive magnetic transitions, which were observed by specific-heat and 
magnetic-susceptibility measurements, were not clearly observed in NQR~\cite{Kam2}. 
The NQR data suggested that the wave vector of the magnetic structure below $T_{\rm N1}$ 
was ($\frac{1}{2}$, $\frac{1}{2}$, $h$) with an incommensurate $k_{z}$ component of the wave vector. 
The magnetic moment at the Ce(2) site has a main component along the crystal $c$ axis and a tilted component within the $ab$ plane. 
This magnetic structure is similar to that of CeRhIn$_{5}$, where the wave vector is expressed with ($\frac{1}{2}$, $\frac{1}{2}$, 0.297)~\cite{Cur1,Bao2}. 
Moreover, taking into account the decrease in signal intensity due to the internal magnetic field at each In site, 
Kambe \textit{et al.} concluded that the localized moment of the Ce(2) ion gives the main contribution to the magnetically ordered state in Ce$_{3}M$In$_{11}$, as suggested first in 
Ref.~\cite{Cus2}. 

In this paper, we performed detailed measurements of the NQR spectra and the spin-lattice relaxation rate $1/T_{1}$ of Ce$_{3}$PtIn$_{11}$ 
with a focus on the In(2), In(3), and In(4) sites. 
In both physical quantities, two successive transitions were clearly observed, which were not detected in the previous study~\cite{Kam1}. 
From the viewpoint of the NQR spectra, each inequivalent In site has a single internal magnetic field between $T_{\rm N1}$ and $T_{\rm N2}$, 
and the In(3) site feels two or more different internal magnetic fields below $T_{\rm N2}$.
Based on the results of our study, we suggest the most probable magnetically ordered state in Ce$_{3}$PtIn$_{11}$, in which  
a wave vector with ($\frac{1}{2}$, $\frac{1}{2}$, 0, or $\frac{1}{2}$) is realized between $T_{\rm N1}$ and $T_{\rm N2}$. 
Our analysis suggests that this wave vector changes into a commensurate or an incommensurate vector depending on the inequivalent Ce sites below $T_{\rm N2}$.

\section{Experiments}

A polycrystalline sample of Ce$_{3}$PtIn$_{11}$ was obtained in a two-step procedure. 
First, CeIn$_2$ and Pt$_3$In$_7$ binaries were synthesized by melting the high-purity elemental constituents under a high-purity argon atmosphere 
using an arc furnace installed inside an MBRAUN glove box with controlled oxygen and moisture contents. 
Then, the two precursors were arc-melted in the same furnace together with elemental indium in the proportion 9:1:8, required to get the nominal composition Ce$_{3}$PtIn$_{11}$. 
The button was turned over and remelted several times to promote homogeneity. 
The final mass loss was less than 1\%. 
The so-obtained sample was wrapped with tantalum foil and annealed in an evacuated quartz ampoule at 500$^{\circ}$C for 4 weeks, 
followed by quenching in cold water.
Quality of the product was checked by x-ray powder diffraction, by energy-dispersive x-ray analysis, and by bulk property measurements. 
the detailed synthesis process and preliminary characterization data are reported elsewhere~\cite{Kac6}. 
The sample contains a partial parasitic phase of CeIn$_{3}$ (less than 5\%-10\%) but has a fairly sharp superconducting transition at $T_{\rm c}\simeq$~0.33~K 
and two antiferromagnetic transitions at $T_{\rm N1}$ = 2.2~K and $T_{\rm N2}$ = 2.0~K, in agreement with the literature data~\cite{Cus1,Kac5}. 
For NQR measurements, the sample was fine powdered in order to reduce the heating-up effect at low temperatures and improve the NQR signal intensity. 
The $^{115}$In NQR studies were performed in the frequency range of 7-70~MHz using a phase-coherent pulsed NQR spectrometer. 
Measurements down to 1.4~K were carried out using a $^{4}$He cryostat. 
The spin-lattice relaxation time $T_{1}$ was obtained from the recovery of the nuclear magnetization after a saturation pulse. 

\section{Results and Discussion}

\subsection{In-NQR spectrum at 4.2~K}

\begin{figure}
\includegraphics[width=8.5cm]{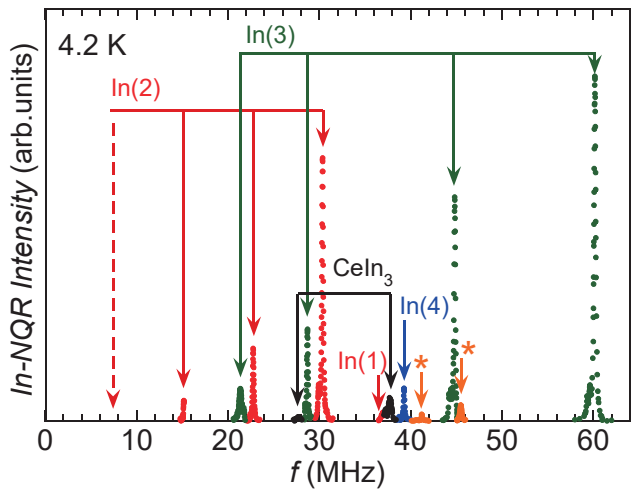}
\caption{
In-NQR spectrum of Ce$_{3}$PtIn$_{11}$ at 4.2 K. 
Signal indicated by * is of parasitic phase. 
}
\label{f2}
\end{figure}

Figure~\ref{f2} shows the NQR spectrum of Ce$_{3}$PtIn$_{11}$ taken at 4.2~K. 
The obtained data are very similar to those reported before~\cite{Kam1,Kam2}.  
As can be inferred from Fig.~\ref{f1}, 
there are four inequivalent crystallographic positions for indium atoms in the unit cell of Ce$_{3}$PtIn$_{11}$, 
namely, In(1) ($4/mmm$), In(2) ($4mm$), In(3) ($2mm$), and In(4) ($2mm$). 
(We follow the notation of site index indicated by Kambe {\it et al.}~\cite{Kam1,Kam2}.) 
For the In nuclei ($I$ = 9/2), the electric quadrupole Hamiltonian $\mathcal{H}_{Q}$ is given by
\begin{equation}
\mathcal{H}_{Q} = \frac{e^{2}qQ}{4I(2I-1)}\Bigl\{ 3I_{z}^{2}-I(I+1)+\frac{\eta}{2}(I_{+}^{2}+I_{-}^{2}) \Bigl\},
\end{equation}
where $eq$, $eQ$, and $\eta$ represent the electric field gradient (EFG), the nuclear quadrupole moment, and the asymmetry parameter of the EFG, respectively. 
$\eta$ is defined by $\frac{V_{xx}-V_{yy}}{V_{zz}}$, where $V_{xx}$, $V_{yy}$, and $V_{zz}$ are the principal axes of the EFG and $|V_{zz}|\geq|V_{yy}|\geq|V_{xx}|$. 
By diagonalizing $\mathcal{H}_{Q}$ and considering the four inequivalent In sites in Ce$_{3}$PtIn$_{11}$, one can assign all the observed features in the NQR spectrum. 
For each In site, four resonance lines appear with increasing frequencies: 
$\nu_{1}$, $\nu_{2}$, $\nu_{3}$, and $\nu_{4}$. 
The $\nu_{1}$ line corresponds to the transition between $|I_{z}=\pm 1/2\rangle$ and $|I_{z}=\pm 3/2\rangle$, the $\nu_{2}$ line is 
due to the transition between $|I_{z}=\pm 3/2\rangle$ and $|I_{z}=\pm 5/2\rangle$, etc. 
Note that $\nu_{1}$ differs from the resonance frequency $\nu_{Q}\equiv\frac{3e^{2}qQ}{2I(2I-1)}$ except for the case of $\eta = 0$. 
For sites with $\eta =0$, they are equidistant, and for those with $\eta \neq 0$, they are not equidistant. 
Therefore, for the In(2) [or In(1)] site, it can be determined as a site with $\eta = 0$ without numerical calculation. 
However, for other sites, it is necessary to perform numerical calculation to identify the site.

The obtained NQR frequency $\nu_{Q}$ and the parameter $\eta$ are 7.591(2)~MHz and 0.000(1), respectively, for In(2), and 15.114(4)~MHz and 0.2388(4), respectively, for In(3). 
We also observed the NQR signals at the positions that were assigned before (see above) as $\nu_{4}$ lines of In(1) and In(4)~\cite{Kam1,Kam2}.
Therefore the NQR spectrum obtained at 4.2 K is consistent with that previously obtained at 3.1 K in the paramagnetic state. 

As shown in Fig.~\ref{f2}, some weak NQR signals were attributable to antiferromagnetically-ordered CeIn$_{3}$. 
In addition, we found two extra lines at around 41.2 and 45.5~MHz, which could be assigned to neither Ce$_{3}$PtIn$_{11}$ nor CeIn$_{3}$ 
and hence likely arise due to other parasitic phases, which were also pointed out previously~\cite{Kam1,Kam2}.

\subsection{In-NQR spectra between 1.4 and 4.2~K}

\begin{figure}
\begin{center}
\includegraphics[width=6cm]{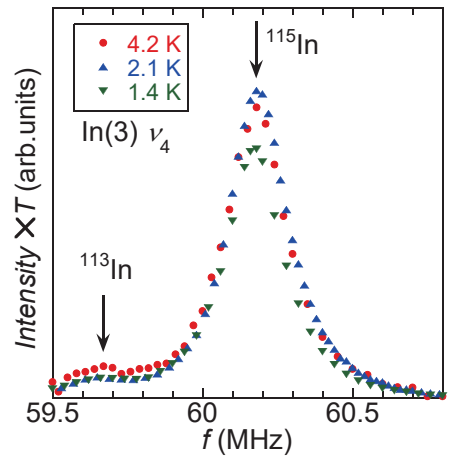}
\caption{
In-NQR spectra of Ce$_{3}$PtIn$_{11}$ for $\nu_{4}$ lines of the In(3) site at 1.4, 2.1, and 4.2~K. 
These spectra are multiplied by temperature and are corrected for the effect of spin-spin relaxation time $T_{2}$. 
}
\label{f3}
\end{center}
\end{figure}

We followed the temperature evolution of the $\nu_{2}$ to $\nu_{4}$ lines of the In(2) site, all lines of the In(3) site, and the $\nu_{4}$ line of the In(4) site. 
In the spectrum recorded at 4.2 K, the $\nu_{4}$ line of the In(1) site 
had quite a short spin-spin relaxation time $T_{2}$. 
If $T_2$ is long enough, 
the NQR signal intensity simply increases in inverse proportion to the temperature, 
but if $T_{2}$ becomes short around $T_{\rm N}$, its correction must be taken into account~\cite{Sli1}. 
Just for this reason, we could not follow  the $\nu_{4}$ line of the In(1) site below $T_{\rm N1}$.
Figure~\ref{f3} shows the spectra of the $\nu_{4}$ line at the In(3) site at 1.8, 2.1, and 4.2~K. 
These spectra were multiplied by temperature and $T_{2}$ corrected. 
The spectral line of  $^{113}$In nuclei is observed on the lower-frequency side of the $^{115}$In line~\cite{Not1}.
Regarding the $\nu_{2}$, $\nu_{3}$, and $\nu_{4}$ lines of the In(2) site as well as the $\nu_{3}$ line of the In(3) site, 
the spectra were almost independent of temperature, similarly to the $\nu_{4}$ line of the In(3) site. 
On the other hand, the $\nu_{1}$ and $\nu_{2}$ lines of the In(3) site and the $\nu_{4}$ line of the In(4) site showed a clear change in the spectrum taken below $T_{\rm N1}$.

\begin{figure}
\begin{center}
\includegraphics[width=7.5cm]{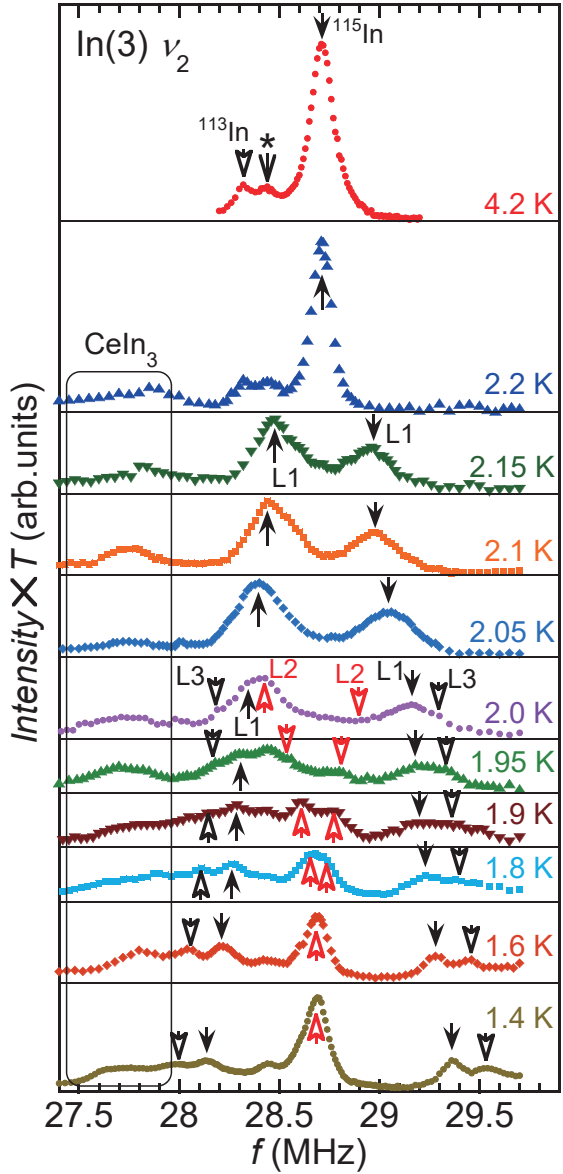}
\caption{
The temperature evolution of the In-NQR spectrum of Ce$_{3}$PtIn$_{11}$ for the $\nu_{2}$ lines of the In(3) site. 
Signal indicated by * is of parasitic phase. 
The characteristic structure observed in the spectra is indicated by several kinds of arrows (L1, L2, and L3). 
}
\label{f4}
\end{center}
\end{figure}

Figure~\ref{f4} shows the temperature evolution of the $\nu_{2}$ line of the In(3) site. 
These spectra were also multiplied by temperature and $T_{2}$ corrected. 
At 4.2~K, we observed a narrow NQR line. 
In addition, a signal from a parasitic phase, as small as that of $^{113}$In nuclei, was 
also observed on the lower-frequency side of the $^{115}$In line~\cite{Not1}.
Notably, with decreasing temperature, the feature splits into two lines in the temperature region between 2.0 and 2.2~K (labeled L1 in Fig.~\ref{f4}). 
The spectra below 2.0~K are more complicated. 
Around the center of the spectra, the separation of lines decreases with decreasing temperature, and the two lines merge into a single line below 1.8~K  (L2 in Fig.~\ref{f4}). 
Below 2.0~K, another couple of separated lines appear outside the original separated lines (L3 in Fig.~\ref{f4}). 
The spectrum changes continuously through $T=T_{\rm N2}$. 
Therefore the phase transition at $T_{\rm N2}$ cannot be interpreted as a phase separation of the paramagnetic phase and the antiferromagnetically ordered phase. 

\begin{figure}
\begin{center}
\includegraphics[width=7.5cm]{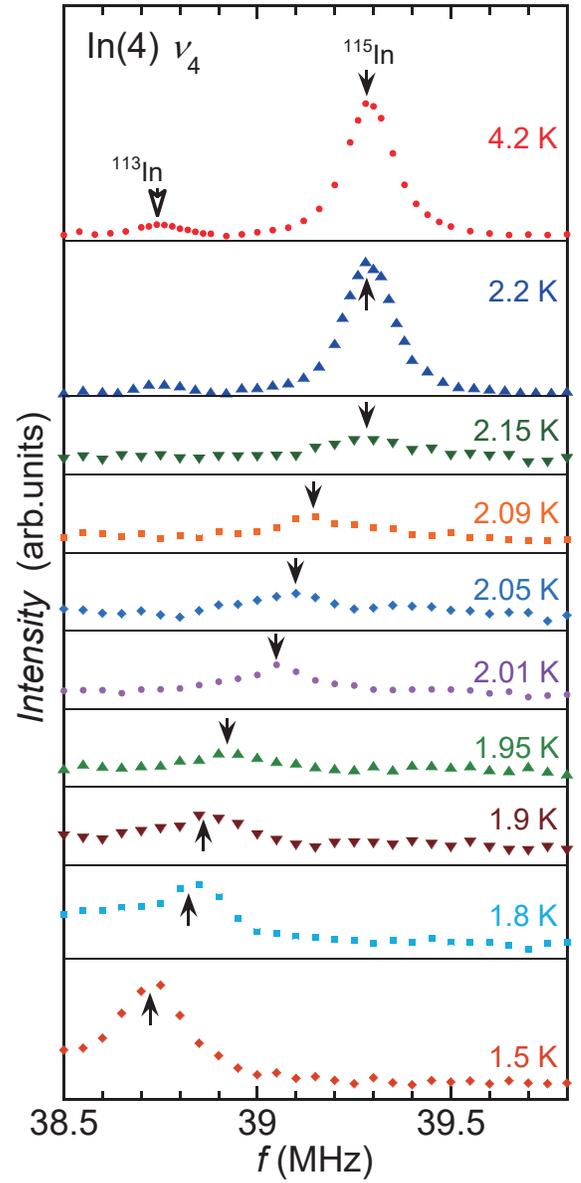}
\caption{
The temperature evolution of the In-NQR spectrum of Ce$_{3}$PtIn$_{11}$ for the $\nu_{4}$ line of the In(4) site. 
The effect of $T_{2}$ was taken into account only for the data obtained at 2.2 and 4.2~K. 
}
\label{f5}
\end{center}
\end{figure}

Figure~\ref{f5} shows the temperature evolution of the $\nu_{4}$ spectrum of the In(4) site. 
The experimental results were $T_{2}$ corrected only for the data obtained at 2.2 and 4.2~K. 
This is because the signal intensity decreased considerably below $T_{\rm N1}$, which hampered precise evaluation of $T_{2}$. 
The recovery of the signal at 1.5~K is probably due to increase of $T_{2}$ below $T_{\rm N2}$. 
Unlike lines of the In(3) site, the $\nu_{4}$ line of the In(4) site shifts monotonically to the low-frequency side without any splitting below $T_{\rm N1}$. 
The peak of the spectrum at 1.5~K is around 38.7~MHz, which is consistent with the previous reports~\cite{Kam1,Kam2}. 

In order to determine the direction of the internal magnetic field and evaluate the contribution of the internal magnetic field at the In(2), In(3), and In(4) sites, 
we added to the Hamiltonian described in Eq.~(1) the following Zeeman-effect Hamiltonian: 
\begin{eqnarray}
\mathcal{H}_{Z} &=& -\mu_{\rm N}\vec{I}\cdot\vec{B}_{\rm int}  \\
&=& -\mu_{\rm N}B_{\rm int}(I_{x}\sin\theta\cos\varphi +I_{y}\sin\theta\sin\varphi +I_{z}\cos\theta ),  \nonumber
\end{eqnarray}
where $\mu_{\rm N}$, $B_{\rm int}$, $\theta$, and $\varphi$ are the nuclear magneton of indium, the internal magnetic field at the In sites, 
the polar angle from $V_{zz}$, and the azimuthal angle from $V_{xx}$, respectively.

\begin{figure}
\includegraphics[width=8.5cm]{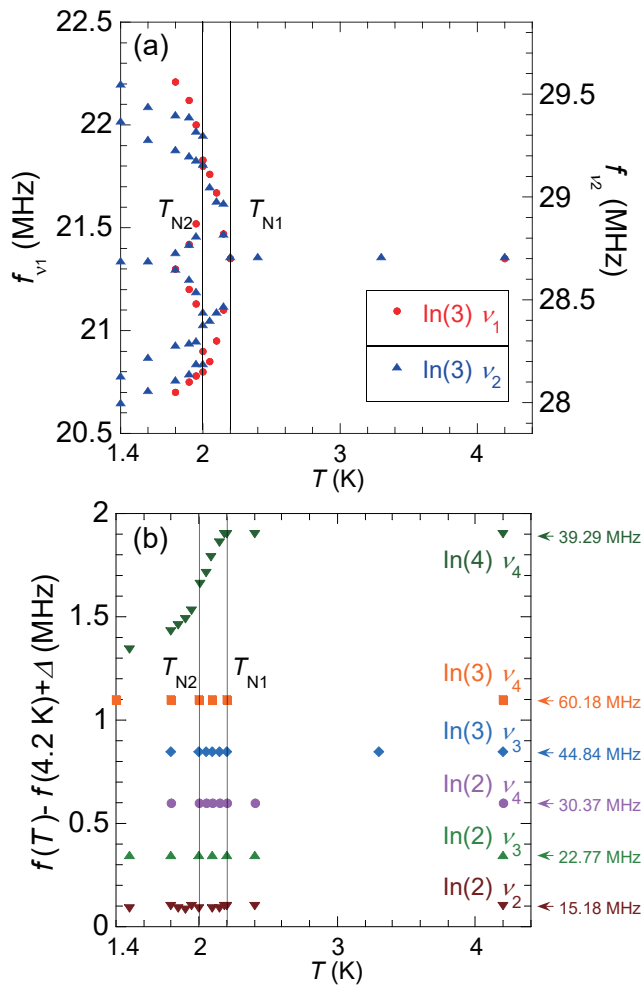}
\caption{
Temperature dependence of the characteristic frequencies in the spectra of Ce$_{3}$PtIn$_{11}$ for (a) the $\nu_{1}$ and $\nu_{2}$ lines of the In(3) site 
and (b) the $\nu_{2}$, $\nu_{3}$, and $\nu_{4}$ lines of the In(2) site, the $\nu_{3}$ and $\nu_{4}$ lines of the In(3) site, and the $\nu_{4}$ line of the In(4) site. 
For the latter case, $f(T)-f(4.2~{\rm K})$ is properly shifted so that the lines do not overlap with each other. 
}
\label{f6}
\end{figure}

Figure~\ref{f6} shows the temperature dependence of the characteristic frequencies 
of each line measured for Ce$_{3}$PtIn$_{11}$. 
First, the $\nu_{2}$, $\nu_{3}$, and $\nu_{4}$ lines of the In(2) site are nearly independent of temperature [see Fig.~\ref{f6}(b)]. 
This feature clearly implies that the internal magnetic field is canceled out at this site. 
Next, it can be seen that both $\nu_{1}$ and $\nu_{2}$ of the In(3) site show almost the same and clear temperature dependences [see Fig.~\ref{f6}(a)]. 
This is in contrast to the fact that both $\nu_{3}$ and $\nu_{4}$ of the In(3) site do not change through out the entire temperature range. 
These results indicate that, at the In(3) site, 
the internal field broadens the $\nu_{1}$ and $\nu_{2}$ lines, yet does not affect the $\nu_{3}$ and $\nu_{4}$ lines.
Performing numerical calculations in terms of Eqs. (1) and (2), one can easily deduce that $\theta = \frac{\pi}{2}$ is realized in the ordered state. 
Moreover, the numerical diagonalization of the Hamiltonians reveals that $\varphi = \frac{\pi}{2}$ is more likely rather than $\varphi = 0$. 
The calculation revealed that a range $\frac{\pi}{3} \lesssim \varphi \leq \frac{\pi}{2}$ is allowed to account for the data 
because of the finite asymmetry parameter $\eta$ at the In(3) site.
Following this finding, the value of the internal magnetic field of 25~mT was found, which can explain the spectral splitting of the In(3) site at $T_{\rm N2}$.
Finally, the peak frequency of the $\nu_{4}$ line of the In(4) site decreases monotonically without splitting below $T_{\rm N1}$. 
This requires at least $\theta = \frac{\pi}{2}$, which implies that the direction of the internal magnetic field at the In(4) site is the same as or close to that at the In(3) site. 
Therefore the internal magnetic field can be evaluated from the shift of the $\nu_{4}$ line of the In(4) site. 
At $T_{\rm N2}$, such a shift attains about 0.3~MHz, which corresponds to an internal magnetic field of about 400~mT.

\subsection{Dipolar model analysis}

Here, we consider the magnetic moments at the Ce sites and the wave vectors, 
which correspond to the obtained direction and magnitude of the internal magnetic fields at $T_{\rm N2}$ 
based on the appearance of the antiferromagnetic order in Ce$_{3}$PtIn$_{11}$. 
We assume the wave vector of ($\frac{1}{2}$,$\frac{1}{2}$) in the $k_{x}$-$k_{y}$ plane of the wave vector space; by analogy to other  
Ce$_{n}M_{m}$In$_{3n+2m}$ compounds, $k =$ ($\frac{1}{2}$, $\frac{1}{2}$, 0) for Ce$_{2}$RhIn$_{8}$, 
($\frac{1}{2}$, $\frac{1}{2}$, $\frac{1}{2}$) for CeIn$_{3}$~\cite{Bao1} and CePt$_{2}$In$_{7}$~\cite{Kur1,Sak1,Rab1,Gau1}, 
and ($\frac{1}{2}$, $\frac{1}{2}$, 0.297) for CeRhIn$_{5}$~\cite{Cur1,Bao2}. 
Most importantly, a similar wave vector ($\frac{1}{2}$, $\frac{1}{2}$, $h$) was recently also suggested for Ce$_{3}$PtIn$_{11}$~\cite{Kam2}.  
Based on the latter finding, we set the moment at the Ce (2) site as $\vec{m} = (m_{\rm a}, m_{\rm b}, m_{\rm c})$ 
and assume that the Ce(1) site is nonmagnetic or very weakly magnetic. 
Then, the internal magnetic fields at the In(1) and In(2) sites are roughly proportional to $(m_{\rm b}, m_{\rm a}, 0)$, 
while those at the In(3) and In(4) sites are proportional to $(0, m_{\rm c}, m_{\rm b})$, if one of the principal axes of the EFG ($V_{zz}$) is 
parallel to the crystal $a$ axis for which the In(3) and In(4) sites are located in the crystal $bc$ plane 
[or $(m_{\rm c}, 0, m_{\rm a})$ if $V_{zz}$ is parallel to the $b$ axis for which the In(3) and In(4) sites are located in the $ac$ plane]. 


First, since the internal magnetic field does not appear at the In(2) site below $T_{\rm N1}$, the ordered magnetic moment has only a $c$-axis component ($m_{\rm a}= m_{\rm b}=0)$. 
Kambe {\it et al}.~\cite{Kam1,Kam2} pointed out that the disappearance of the signal at the In(1) site implies the appearance of an internal magnetic field 
parallel to the principal axis $V_{zz}$ of the EFG. 
In such a case, a similar clear internal magnetic field would have appeared at the In(2) site, and all the In(2) lines must split, which is inconsistent with our experimental results. 
Therefore it is reasonable to presume that the extremely short relaxation time $T_{2}$ makes difficult observation of the signal at the In(1) site  below $T_{\rm N1}$. 
In addition, Kambe {\it et al.}~\cite{Kam1,Kam2} suggested some possibility of a wave vector of (0, 0, $\frac{1}{2}$) with a magnetic moment direction along the $c$ axis. 
However, in such a case, a large internal magnetic field would have also appeared at the In(2) site. 
Hence this conjecture must be ruled out.

 \begin{table}[t]
  \caption{$\nu_{Q}$(calc) and $\eta$(calc) are calculated based on the electronic structure calculation using the WIEN2K code~\cite{Bla1}. 
  $\nu_{Q}$(exp) and $\eta$(exp) are experimentally obtained at 4.2~K.}
  \label{t1}
  \begin{center}
   \begin{tabular}{@{\hspace{\tabcolsep}\extracolsep{\fill}}ccccc} \hline\hline
    & $\nu_{Q}$(calc) (MHz) & $\eta$(calc)  & $\nu_{Q}$(exp) (MHz) & $\eta$(exp) \\ \hline
    &&&& \\
   In(1) & 10.1  & 0 & (9.24) & \\
    &&&& \\
   In(2) & 8.73  & 0 & 7.591(2) & 0.000(1) \\ 
    &&&& \\ 
   In(3) & 14.5  & 0.190 & 15.114(4) & 0.2388(4) \\
    &&&& \\
   In(4) & 10.4  & 0.00054 & ($\sim$9.8) & \\ 
    &&&& \\ \hline\hline
   \end{tabular}
  \end{center}
 \end{table}

Next, in the case that $V_{zz}$ is parallel to the crystal $a$ axis [the In(3) sites located in the $bc$ plane], 
an internal magnetic field parallel to (0, $m_{\rm c}$, 0) should appear at the In(3) site 
because the in-plane component of the magnetic moment must be zero ($m_{\rm a}=m_{\rm b}=0$: see the discussion above). 
In order to confirm the relation between the principal axes of the EFG and the crystal axes, 
we performed the electronic structure calculation using the WIEN2K code~\cite{Bla1}, 
where the full-potential linearized augmented-plane-wave method with the generalized gradient approximation for electron correlations is used.
For calculations, the space group, the lattice constants, and the
starting positions of atoms in the cell were taken from Tables~\ref{t1} and \ref{t2} of Ref.~\cite{Kac5}, 
and the positions of the atoms were finally determined by structure optimizations. 
For comparison, Table~\ref{t1} summarizes the calculated and the experimentally determined $\nu_{Q}$ and $\eta$. 
Both values are consistent with each other, which ensures the reliability of the calculations. 
The calculated values of $V_{aa}$, $V_{bb}$, and $V_{cc}$, before the rotation of the principal axes, are 17.75, -10.56, and -7.19, respectively in units of $10^{21}$~V/m$^{2}$. 
These values were calculated for the In(3) sites located in the crystal $bc$ plane. 
Hence this means that $V_{xx}$, $V_{yy}$, and $V_{zz}$, after the rotation of the principal axes, are parallel to the $c$ axis, $b$ axis and $a$ axis, respectively, which
implies that the internal magnetic field at the In(3) site is perpendicular to both $V_{zz}$ and $V_{xx}$. 
This is consistent with the experimentally obtained direction of the internal magnetic field at the In(3) site 
($\theta = \frac{\pi}{2}$, $\frac{\pi}{3} \lesssim \varphi \leq \frac{\pi}{2}$ ).
At least between $T_{\rm N1}$ and $T_{\rm N2}$, the In(3) site feels a homogeneous internal magnetic field of $\simeq 25$~mT. 
Otherwise, the spectrum would be much broader and the obtained values of the internal magnetic fields would also be distributed.
This adds a condition that the wave number along the $k_{z}$ direction must be commensurate. 
Accordingly, the wave vector should be either ($\frac{1}{2}$, $\frac{1}{2}$, 0) or ($\frac{1}{2}$, $\frac{1}{2}$, $\frac{1}{2}$), 
and the magnetic moment must be parallel to the $c$ axis between $T_{\rm N1}$ and $T_{\rm N2}$. 
This conclusion is different from the suggestion by the previous report that required inclination of the magnetic moment from the $c$ axis~\cite{Kam2}.

 \begin{table}[t]
  \caption{Antiferromagnetic wave vectors, magnetic moments, and the dipole field sum at the In(3) and In(4) sites 
           for the experimental internal magnetic fields of the In(3) and In(4) sites at $T_{\rm N2}$.
           Calculations were performed for the In(3) and In(4) sites located in the crystal $bc$ plane. 
           }
  \label{t2}
  \begin{center}
   \begin{tabular}{@{\hspace{\tabcolsep}\extracolsep{\fill}}ccccc} \hline\hline
    & Wave vector & $m/\mu_{\rm B}$  & $\vec{B}_{\rm In(3)}$ (mT) & $\vec{B}_{\rm In(4)}$ (mT) \\ \hline
    &&&& \\
   Ce(1) & ($\frac{1}{2}$, $\frac{1}{2}$, 0)  & 0.20 & (0, 15.4, 0) & (0, -15.4, 0) \\
    &&&& \\
   Ce(2) & ($\frac{1}{2}$, $\frac{1}{2}$, 0)  & 5.30 & (0, 9.2, 0) & (0, 416.7, 0) \\ 
    &&&& \\
   (Total) &&& (0, 24.6, 0) & (0, 401.3, 0) \\ 
    &&&& \\ \hline
    &&&& \\
   Ce(1) & ($\frac{1}{2}$, $\frac{1}{2}$, $\frac{1}{2}$)  & 0.19 & (0, 15.3, 0) & (0, -14.6, 0) \\
    &&&& \\
   Ce(2) & ($\frac{1}{2}$, $\frac{1}{2}$, $\frac{1}{2}$)  & 5.29 & (0, 9.6, 0) & (0, 415.9, 0) \\ 
    &&&& \\
   (Total) &&& (0, 24.9, 0) & (0, 401.3, 0) \\ 
    &&&& \\ \hline\hline
   \end{tabular}
  \end{center}
 \end{table}

By calculating the dipole magnetic field with ($\frac{1}{2}$, $\frac{1}{2}$, 0) or ($\frac{1}{2}$, $\frac{1}{2}$, $\frac{1}{2}$), 
we have checked whether the experimentally obtained internal magnetic fields at $T_{\rm N2}$ could be reproduced, i.e.,  
25~mT at the In(3) site and 400 mT at the In(4) site. 
For the numerical calculations of the dipole field with the magnetic moment along the crystal $c$ axis, 
120$\times$120$\times$30 unit cells were used, and it was confirmed that the calculations converged by the calculation radius of about 10~nm. 
Here, we used the structure data (space group, lattice constants, and positions of atoms) summarized in Tables~\ref{t1} and \ref{t2} of Ref.~\cite{Kac5}. 
First, we assumed the magnetic moment only at the Ce(2) site. 
However, in this case, we were not able to determine the size of the magnetic moment that could explain the internal magnetic fields at both In sites~\cite{Not2}. 
Next, we additionally assumed the magnetic moment at the Ce(1) site. 
For simplification, we set the magnetic moment of the Ce(1) site parallel to that of the Ce(2) site belonging to the same unit cell.  
Then, we obtained the set of the magnetic moments summarized in Table~\ref{t2}, which is consistent with the experimentally obtained internal magnetic fields. 
The contribution of the internal magnetic field at the In(3) site is attributable to the magnetic moments of the Ce(1) and Ce(2) sites. 
On the other hand, the main contribution of the field at the In(4) site is ascribable to the moment of the Ce(2) site. 
Remarkably, the moment of the Ce(1) site is about 4\% of that of the Ce(2) site in both cases. 
This indicates that the magnetic moment of the Ce(1) site is very small and 
that the magnetic moment of the Ce(2) site plays a main role in the magnetism of Ce$_{3}$PtIn$_{11}$. 
The evaluated moment for the Ce(2) site is much larger than $g_J = 15/7 \simeq 2.14~\mu_{\rm B}$ expected for the Ce$^{3+}$ ion. 
In the actual system, each In site feels a pseudo-dipole field transferred by conduction electrons~\cite{Cur2},
which is larger than a direct dipole field. 
In the present simple calculation, this contribution of the transferred field cannot be included. 

Below $T_{\rm N2}$, about half the In(3) sites do not feel any internal magnetic fields, 
while the In(4) sites feel nearly identical magnetic field. 
If the In(4) site exhibits field cancellation, its $\nu_{4}$ line must split below $T_{\rm N2}$.  
It is worth highlighting that the internal magnetic field of the In(3) site is mainly affected by the moment contribution from the Ce(1) site 
and some partial contribution from the Ce(2) site. 
On the other hand, the internal field at the In(4) site is influenced by the magnetic moments of the  Ce(2) site only. 
This suggests that the wave vector of the Ce(2) site remains commensurate and that of the Ce(1) site changes into incommensurate 
in order to compensate the constant contribution, at the In(3) site, of the Ce(2) moments by modulation of the Ce(1) moments. 
If we do not consider the moment of the Ce(1) site, the commensurate order of the moment of the Ce(2) site produces 
double-peak structure for the $\nu_1$ and $\nu_2$ lines of the In(3) sites, even below $T_{\rm N2}$, which was observed between $T_{\rm N1}$ and $T_{\rm N2}$. 
This is inconsistent with the experimentally observed center line and broad satellite lines. 
Though most of the Ce(1) moment seems to have been screened by the Kondo effect, 
the existence of the tiny magnetic moment of the Ce(1) site is essential to explain the spectrum below $T_{\rm N2}$. 
Since the transition at $T_{\rm N2}$ is clearly observed in thermodynamic quantities such as specific heat~\cite{Cus1,Cus2,Kac5} and $1/T_{1}$ (described in the next section), 
the large magnetic moment of the Ce(2) site is responsible for the transition. 
This implies that the change of the wave vector at the Ce(2) site is essential, yielding a commensurate-to-commensurate transition: 
($\frac{1}{2}$, $\frac{1}{2}$, 0) to ($\frac{1}{2}$, $\frac{1}{2}$, $\frac{1}{2}$) or vice versa. 
Hence a wave vector of ($\frac{1}{2}$, $\frac{1}{2}$, $h$) at the Ce(1) site with modulation in the $k_{z}$ direction 
from ($\frac{1}{2}$, $\frac{1}{2}$, 0) or ($\frac{1}{2}$, $\frac{1}{2}$, $\frac{1}{2}$) 
may account for the absence of any internal magnetic field at the In(3) site. 
This model was suggested before for the magnetic moment of the Ce(2) site by Kambe \textit{et al.}~\cite{Kam2}. 
However, in this case, the spectrum should be broadened for general $h$. 
Probably, $h$ is modulated in the $k_{z}$ direction in a way that brings about the cancellation of the internal magnetic field at half of the In(3) sites, 
although the exact $h$ could not be determined within the present calculations. 
Further analysis beyond the simple dipole sum calculations is 
desired.

\subsection{Spin-lattice relaxation rate $1/T_{1}$}

\begin{figure}
\includegraphics[width=8.5cm]{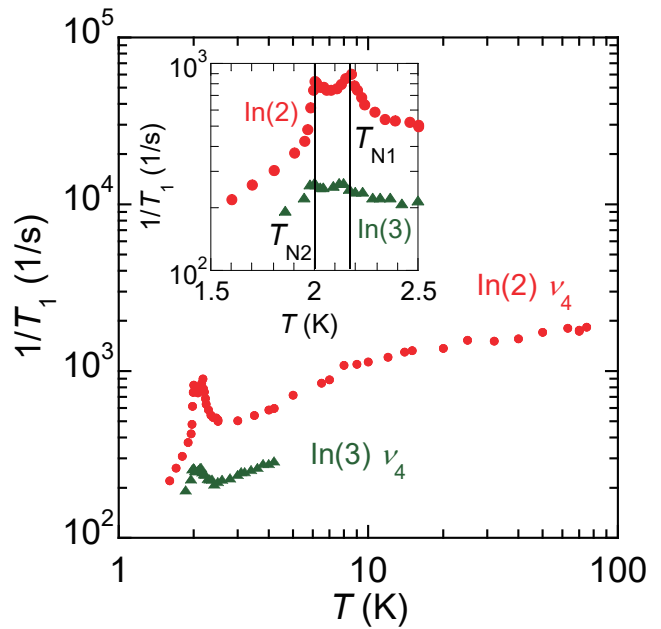}
\caption{
Temperature dependence of the spin-lattice relaxation rate $1/T_{1}$ of Ce$_{3}$PtIn$_{11}$ for $\nu_{4}$ lines of the In(2) and In(3) sites. 
Inset: magnified temperature dependence of $1/T_{1}$ between 1.5 and 2.5~K.}
\label{f7}
\end{figure}

Figure~\ref{f7} shows the temperature dependence of the spin-lattice relaxation rate $1/T_{1}$ of the $\nu_{4}$ line at the In(2) and In(3) sites. 
In both cases, two distinct transitions are observed at $T_{\rm N1} \simeq 2.2$~K and $T_{\rm N2} \simeq 2.0$~K. 
The observed relaxation curve can be fitted well to the theoretical formula expected for the $\nu_{4}$ lines of $I = 9/2$ nuclei~\cite{Chep}. 
This result indicates that the longitudinal relaxation in Ce$_{3}$PtIn$_{11}$ is of magnetic origin and  
that there is almost no influence of the internal magnetic field even below $T_{\rm N1}$, 
which is consistent with the conclusion that the internal field is absent and small at the In(2) and In(3) sites, respectively.
Here, it should be recalled that Kambe {\it et al.} pointed out that superconductivity is observed at the In sites that feel the internal magnetic fields~\cite{Kam2}. 

\section{Summary}
NQR measurements were performed on the heavy-fermion superconductor Ce$_{3}$PtIn$_{11}$. 
The temperature dependencies of both $1/T_{1}$ and the NQR spectra revealed the occurrence of two successive magnetic transitions. 
Each inequivalent In site feels a single internal magnetic field between $T_{\rm N1}$ and $T_{\rm N2}$, 
and the In(3) site feels two or three different internal magnetic fields below $T_{\rm N2}$. 
Based on the NQR data we concluded that the magnetic wave vector is commensurate ($\frac{1}{2}$, $\frac{1}{2}$, 0, or $\frac{1}{2}$) 
with the magnetic moment parallel to the $c$ axis between $T_{\rm N1}$ and $T_{\rm N2}$. 
It was also revealed that the ordered magnetic moment at the Ce(2) site is about 25 times larger than that at the Ce(1) site. 
This indicates that the magnetic moment of the Ce(2) site plays a crucial role in the magnetism of Ce$_{3}$PtIn$_{11}$. 
The duality of the crystallographically inequivalent Ce sites is indeed realized in this compound. 
At $T_{\rm N2}$, a complicated phase transition of commensurate to commensurate at the Ce(2) site and commensurate to incommensurate at the Ce(1) site is suggested. 
Further analysis beyond the simple dipolar model presented in this paper is desired. 
NQR studies performed at lower temperatures are required to conclude 
that the claimed coexistence of antiferromagnetism and superconductivity in Ce$_{3}$PtIn$_{11}$ is an intrinsic property of this interesting material.


\vspace{5mm}
\begin{center}

{\bf ACKNOWLEDGMENTS}

\end{center}

\vspace{5mm}


The authors thank S. Kambe and T. Ohama for fruitful discussions. 
This work was supported by JSPS KAKENHI Grant (No. 18K03505 and No. 19K14644).
The work in Poland was supported by the National Science Centre (Poland) under Research Grant No. 2015/19/B/ST3/03158.


\begin{thebibliography}{99}
\bibitem{Not1} 
Because the natural abundance of $^{113}$In is 4.3\% and that of $^{115}$In is 95.7\%,
the $^{113}$In line intensities are quite low (see Fig.~\ref{f3}). 
The lines due to $^{113}$In overlap with those of $^{115}$In because the ratio of quadrupole moments is $^{113}Q/^{115}Q = 0.759/0.770$. 
However, they can be identified because of this ratio and the natural-abundance ratio.
\bibitem{Not2} 
Strictly speaking, it is not impossible to explain the spectrum between $T_{\rm N1}$ and $T_{\rm N2}$ 
solely by the moment of the Ce(2) site if we consider transferred magnetic fields described below, which may be different at the In(3) and In(4) sites. 
However, it becomes more difficult to explain the spectrum below $T_{\rm N2}$ without the moment of the Ce(1) site.
\end{thebibliography}
\end{document}